\documentstyle[aps,preprint,amsfonts,epsf]{revtex}
\begin{document}
\preprint{LA-UR-93-3621}

\title{
Impurities and Quasi-One Dimensional Transport\\ in a
$\protect\bbox{D}$-wave Superconductor}
\author{
A.\ V.\ Balatsky$^1$\protect\cite{aa}, A.\ Rosengren$^2$, B.\ L.\ Altshuler$^3$
}
\address{
$^1$ Theoretical Division, Los Alamos National Laboratory, Los Alamos, NM
87545\\
$^2$ Superconductivity Technology Center, Los Alamos National Laboratory,
NM 87545 and\\ Department of Theoretical Physics, Royal Institute of
Technology, S-10044, Stockholm, Sweden\protect\cite{ab}\\
$^3$ Department of Physics, MIT, Cambridge, MA 02139 }
\date{October 28, 1993}
\maketitle

\begin{abstract}
Impurity scattering in the unitary limit produces low energy
quasiparticles with anisotropic spectrum in a two-dimensional $d$-wave
superconductor. We describe a new {\em quasi-one-dimensional } limit of
the quasiparticle scattering, which might occur in a superconductor
with short coherence length and with {\em finite} impurity potential
range. The  dc  conductivity in a $d$-wave superconductor is
predicted to be proportional to the normal state scattering rate and is
impurity-{\em dependent}. We show that the {\em quasi-one-dimensional }
regime occurs in high-$T_c$ superconductors at low temperatures
$T\lesssim 10$~K. We argue that the impurities produce weak
localization, in the {\em orthogonal } universality class, of the
quasiparticles in a $d$-wave superconductor.

\end{abstract}



The symmetry of the pairing state in high-$T_c$ superconductors was
addressed in some recent experiments on the $T^3$ dependence of the
NMR~\cite{Slichter}, phase shift $\pi$ in the flux dependence of the
Josephson current~\cite{Tony}, linear temperature dependence of the
penetration depth $\lambda(T)$ in single crystals~\cite{Hardy}, and
strong anisotropy of the energy gap in  angular resolved
photoemission~\cite{Shen}. All of them support $d_{x^2-y^2}$
symmetry of the gap. Theoretically the $d$-wave pairing state is
predicted in spin-fluctuation exchange models~\cite{MBP,Ueda} as well
as in models with strong correlations~\cite{Doug}. On the other hand
there is a body of experimental results which seems to be in accord
with  $s$-wave pairing, such as the gapless $s$-wave state, found
in~\cite{Aeppli}, and the BCS-like behavior of the penetration depth in
electron-doped system~\cite{Wu}.

It is well known that scalar impurities are pair breakers in
$d$-wave and any other nontrivial pairing state
superconductor~\cite{Rice,Gorkov,Lee}. They produce a finite lifetime
of the quasiparticles in the nodes of the gap, a finite density of
states at low energy, and a {\em finite} low frequency conductivity at
low temperatures, ignoring localization effects. For the special case
of a two-dimensional superconductor with a $d$-wave gap, a
straightforward calculation yields the surprising result that the {\em
dc} conductivity $\sigma(\omega\to 0)$ is a ``universal''
number~\cite{Lee}, {\em independent} of the lifetime of quasiparticle
(but dependent on the anisotropy ratio of the velocities of the
quasiparticle in the node of the gap)~\cite{Fradkin}. However, experiments on
microwave absorption in YBCO crystals with Zn impurities~\cite{Bonn}
show a {\em linear} temperature dependence of the conductivity and an
impurity-{\em dependent} low-temperature conductivity. The matter is
further complicated by the fact that the data in~\cite{Bonn} also
exhibit no saturation  at low temperature. This as
well might suggest that the conductivity is not impurity dominated in
this temperature range.

The purpose of this letter is to address {\em the role of a strongly
scattering disorder with finite impurity range on the dc conductivity}
at low temperatures in a short coherence length superconductor. (i)~We
will show that there is a new {\em quasi-one-dimensional regime} for
the  dc  conductivity in superconductors with a short coherence
length $\xi$, comparable to the range of the impurity potential
$\lambda$. The quasiparticle contribution to the {\em dc } conductivity
is governed by the self-energy $\Sigma(\omega\to0)=-i\gamma$ and by the
phase space available for low-energy quasiparticles. The quasiparticle
dispersion is strongly anisotropic in the vicinity of the nodes in a
$d$-wave superconductor that has $E_k=\sqrt{v_1^2k_1^2+v_F^2k_3^2}$
and $v_1/v_F\sim\Delta_0/\epsilon_F$ (see Fig.~\ref{fig1}). We find
that the overall contribution to the conductivity depends on the ratio
of the energy of the quasiparticle to the scattering rate
$v_1\lambda^{-1}/\gamma$, and we get at $T=0$:
\begin{equation}
\sigma(\omega\to 0)=\frac{e^2}{2\pi\hbar}~\frac{2}{\pi^2}~
\frac{v_F}{v_1}~\left(1+\left(\frac{\gamma}
{2v_1\lambda^{-1}}\right)^2\right)^{-1/2}.
\label{sigv}
\end{equation}
For $v_1\lambda^{-1}/\gamma\ll 1$ the quasiparticle dynamics is
essentially {\em quasi-one-dimensional } and conductivity {\em depends}
on the impurity concentration. Our model predicts that the  dc
conductivity at low temperature should be {\em proportional to the
scattering rate in the normal state}. This limit might occur in
high-$T_c$ superconductors, for which we estimate $\lambda/a \sim 1-3$
and $\Delta_0/\epsilon_F\sim 10^{-1}$. In the limit $\lambda\to 0$
Eq.~(\ref{sigv}) gives the ``universal''  dc  conductivity, found
in~\cite{Lee}. (ii)~We argue that the origin of the strong potential
scattering due to impurities in the high-$T_c$ superconductors is the
highly correlated antiferromagnetic nature of the normal state. The
range of the impurity potential might be of the order of $\xi_{AFM}$
and thus comparable to the superconducting coherence length $\xi$.
Under these assumptions retaining a {\em finite} range of the impurity
potential is required. (iii)~We also discuss the localization of
quasiparticles close to the nodes, even for a slowly varying random
potential in a $d$-wave superconductor with scalar impurities. We argue
that scalar disorder leads to a weak {\em orthogonal } localization of
quasiparticle states \cite{Lee,Fradkin}. All of the results, presented
here, are valid for {\em any} superconductor with nodes in 2D with a
Dirac spectrum of quasiparticles.

\begin{figure}
\epsfxsize=2in
\centerline{\epsfbox{condfig.ps}}
\caption{
Graphical presentation of the $d_{x^2-y^2}$ state, where
$\Delta(\protect\bbox{k})=\Delta_0(\cos k_xa-\cos k_ya)$. For clarity
the gap function is only drawn in the neighborhood of one node
$\protect\bbox{k}_0$. For calculational purposes a coordinate system
$(k_1,~k_3)$ with the origin at $\protect\bbox{k}_0$ is used instead of
$(k_x,~k_y)$ with the origin at the center of the Brillouin zone
$\Gamma$. They are related by $k_1=(k_x-k_y)/\protect\sqrt{2},~k_3=
(k_x+k_y)/\protect\sqrt{2}-|\protect\bbox{k}_0|$. The  FS denotes the Fermi
surface of the tight-binding band $\xi(\protect\bbox{k})=-t(\cos
k_xa+\cos k_ya)-\mu$. After linearization around $\protect\bbox{k}_0$
we find $\Delta(\protect\bbox{k})=v_1k_1,~\xi(\protect\bbox{k})=
v_Fk_3$, where $v_1=-\protect\sqrt{2}\Delta_0
a\sin(k_{0x}a) and ~v_F=-\protect\sqrt{2}ta\sin(k_{0x}a)$. The momentum
sums performed will be cut off at $|\protect\bbox{k}|=2/\lambda$, where
$\lambda$ is the range of the impurity potential.\label{fig1}}
\end{figure}

We consider scalar impurities that  give rise to the randomly
distributed strong scatterers in 2D with a {\em finite} range
$\lambda$: $\langle U({\bf r})U(0)\rangle
=(\eta/\lambda^2\pi)\exp(-r^2/\lambda^2)\stackrel{\lambda\to
0}{\longrightarrow}\eta\delta({\bf r})$ and dispersion $\eta$. The
assumption of strong potential scattering off the impurity sites is
well accepted for heavy fermion systems, where the Kondo effect plays
an important role and thus any scalar impurity might produce the
``Kondo hole'' with the $s$-wave scattering phase shift $\delta_0$
close to $\pi/2$. The same assumption for the high-$T_c$
superconductors seems to be justified by the experiments on Zn
impurities in YBCO, which produce gapless superconductivity at the 3\%
doping level and strongly change the NMR linewidth of
$^{63}$Cu~\cite{Walstedt}. The possible model describing  the $Zn$
impurities in high-$T_c$ was proposed recently \cite{MP}.  We therefore
assume that scalar impurities are strong scatterers in these
superconductors.

The second assumption of a {\em finite} range of the impurity potential
is motivated by the observation that the high-$T_c$ superconductors
have a substantial antiferromagnetic coherence length $\xi_{AFM}\sim
3a$ at the transition temperature. Thus a scalar impurity {\em will }
produce distortions in the magnetic correlations on the range of the
$\xi_{AFM}$. On the other hand the superconducting coherence length
$\xi\sim 20$~{\AA} is comparable to this scale and thus, the range of
the potential is finite on the scale relevant for superconductivity.
This point should be contrasted to the case of heavy-fermion
superconductors, where the coherence length is $\sim 10^3$~{\AA}, and
therefore, any potential impurity will have its range substantially
shorter than the coherence length. In that case it is reasonable to
use instead the assumption that the impurity has effectively {\em zero}
range. We will retain $\lambda$ finite below. As we will show, this
leads to the new parameter $v_1\lambda^{-1}/\gamma$ and a {\it dc}
conductivity that is dependent on this parameter.

The Bogoliubov Hamiltonian for quasiparticles in a $d$-wave 2D
superconductor is:
\begin{eqnarray}
H=\int ~d\bbox{r}\Psi^{\dag}(\bbox{r})(\xi(\bbox{r})\tau_3
+\Delta(\bbox{r})\tau_1+U(\bbox{r})\tau_3)\Psi(\bbox{r}),
\label{ham}
\end{eqnarray}
where $\Psi =(c^{}_{\uparrow},~c^{\dag}_{\downarrow})$ is the Nambu
spinor, $\tau_i$ are the Pauli matrices, $\xi(\bbox{k})=-t(\cos
k_xa+\cos k_ya)-\mu$ is the energy, counted from the Fermi surface,
$\Delta(\bbox{k})=\Delta_0(\cos k_xa -\cos k_ya)$ is the $d_{x^2-y^2}$
energy gap, and $U(\bbox{r})$ is the impurity potential. For the low-energy
states, we linearize the Hamiltonian in the vicinity of the node
close to the $(\pi /2,~\pi /2)$ point. We find
$\Delta(\bbox{k})=v_1k_1,~\xi(\bbox{k})=v_Fk_3$ in the new coordinates,
defined in Fig.~\ref{fig1}~\cite{Lee}. The resulting Dirac-like
Hamiltonian takes the form:
\begin{eqnarray}
H=\int d\bbox{r}~\Psi^{\dag}((v_F\hat{k}_3+U(r))\tau_3
+v_1\hat{k}_1\tau_1)\Psi.
\label{Dirac}
\end{eqnarray}
The self-consistent Green functions are
given by
\begin{eqnarray}
&G=-(i{\tilde\omega}_n+\xi(\bbox{k}))/D,~F=-\Delta(\bbox{k})/D,
\label{green}
\end{eqnarray}
with $D={\tilde\omega}_n^2+\xi^2(\bbox{k})+\Delta(\bbox{k})^2
$ ,    $i{\tilde\omega}_n=i\omega_n-\Sigma(i\omega_n)$ and we ignore the
self-energy contribution to the anomalous Green function from the
impurity scattering, a contribution that vanishes upon angular
integration~\cite{gap}. For the strong potential scattering impurities
the normal self-energy is given by~\cite{PP,Hirschfeld}:
\begin{eqnarray}
\Sigma(i\omega_n)= \Gamma g_0(i\omega_n)/\left [c^2-g^2_0(i\omega_n)\right ],
\label{sigma}
\end{eqnarray}
where $\Gamma =n_i/\pi N_0$ is the scattering rate in the normal phase,
$c=\cot\delta_0,~g_0(i\omega_n)=4(\pi
N_0)^{-1}\mathop{{\sum}'}_{\bbox{k}}G(\bbox{k},~i\omega_n)$, the factor
of 4 in front of the last sum reflects the number of nodes in the gap,
$n_i$ is the impurity concentration, and $N_0$ is the density of states
at the Fermi surface. The prime in $\mathop{{\sum}'}$ stands
for the momentum sum up to the cut off $|\bbox{k}|<2\lambda^{-1}$ and
comes from the {\em finite} range of the impurity correlator $\langle
U_{\bbox{p}}U_{-\bbox{p}}\rangle =\eta\exp(-p^2\lambda^2/4)$, which is
implemented as the ``hard'' cut off. This momentum cut off follows
immediately from the derivation of the self-energy with a finite range
of the potential, see for example~\cite{Hirschfeld}. We find from the
solution of the Dyson equation, that $\Sigma(i\omega_n)$ is also
momentum dependent with characteristic range $\lambda^{-1}$, which will
be taken into account in the conductivity calculation.

The Born scattering limit is recovered from the Eq.~(\ref{sigma}) for
$c^2\gg g_0(i\omega_n)$. We are interested in the case of {\em unitary}
scattering, for which we take the $s$-wave scattering amplitude
$\delta_0=\pi /2,~c=0$ for scattering momentum $|q|< \lambda^{-1}$. In this
limit the solution of the
Eq.~(\ref{sigma}) for $\Sigma(\omega\rightarrow0)=-i\gamma$ is
\begin{eqnarray}
\gamma^2 =\frac{\Gamma\pi N_0}{4}~\left[\mathop{{\sum}'}_{\bbox{k}}
\left(\gamma^2+(v_1k_1)^2+(v_Fk_3)^2\right)^{-1}\right]^{-1}.
\label{gamma}
\end{eqnarray}
The sum over momentum in this equation is logarithmically divergent at
the upper limit, which is taken as $\min(\lambda^{-1},~\Delta_0/v_1)$
for $k_1$ integral; the $k_3$ integral is always cut off by
$\Delta_0/v_F$, since
$v_F\lambda^{-1}\sim\epsilon_F\gg\gamma,~\Delta_0$. We find for
$I=\mathop{{\sum}'}_{\bbox{k}}~(\gamma^2+(v_1k_1)^2+(v_Fk_3)^2)^{-1}$
\begin{equation}
I= \ln\left(b+\sqrt{1+b^2}\right)/ \left (2\pi v_1v_F \right ),
\label{I}
\end{equation}
where $b=\min(2v_1\lambda^{-1},~\Delta_0)/\gamma$. $I$ has two
asymptotics:

1) $I=1/(2\pi v_1v_F)(\ln(\Delta_0/\gamma)+O(\gamma/\Delta_0))$, for
$v_1\lambda^{-1}/\gamma\gg 1$. This limit corresponds to the isotropic
strong scattering with no momentum cut off as $\lambda^{-1}\to\infty$.
It has been investigated previously~\cite{Lee,Hirschfeld,Doug2}. The
impurity self-energy is shown to be $\tilde{\gamma}
=\Delta_0\sqrt{{\displaystyle{\pi\Gamma\over{\Delta_0\ln(\Delta_0
/\Gamma)}}}}$. We will not further discuss this case.

2) $I=\lambda^{-1}/(\pi v_F)\gamma +1/(2\pi
v_1v_F)O((2v_1\lambda^{-1}/\gamma)^3)$, for $v_1\lambda^{-1}/\gamma\ll
1$. In this case the lifetime is given by:
\begin{eqnarray}
\gamma =\frac{\Gamma\pi}{8}~p_F\lambda\ .
\label{gamma2}
\end{eqnarray}

To check the self-consistency of the assumption
$v_1\lambda^{-1}/\gamma\ll 1$ we use Eq.~(\ref{gamma2}) to get
$\Gamma(p_F\lambda)^2\gg 8/\pi \Delta_0\sim 8T_c$. For an  estimated
$\lambda\sim 2a$  the condition for the {\em
quasi-one-dimensional } regime of quasiparticle scattering is
\begin{eqnarray}
\Gamma\ge 8\cdot10^{-2} \Delta_0\sim 2.5\cdot10^{-1}T_c\simeq 20\mbox{ K}.
\end{eqnarray}
This estimate is supported by the typical scattering rate in clean
samples of YBCO $\Gamma /T_c\sim 5\cdot 10^{-2}$. Furthermore, in the
impure samples with {\em quadratic} temperature dependence of the
penetration depth the estimates are $\Gamma /T_c\sim 1$~\cite{Gold}.
Thus the {\em quasi-one-dimensional } regime of the quasiparticle
scattering should occur at low temperatures $T<\gamma$ in not too clean
samples.

To explain this effective {\em change of dimensionality} we note that
the transverse momentum in the sum in case 2) is limited by
$k_1<2/\lambda$ and the quasiparticle dispersion on such a small scale
is irrelevant, compared to $\gamma$. The transverse scattering does not
contribute effectively to the conductivity; we call this case a {\em
quasi-one-dimensional } limit. The existence of this limit is the
result of the {\em finite} impurity range $\lambda$~\cite{Born}. In
this limit the scattering rate in the superconducting state is of the
same order as the normal state scattering rate
$\gamma\sim 2\Gamma\sim 40$~K for $p_F\lambda\sim 6$ and is
similar  to the scattering rate in the {\em isotropic} limit:
$\gamma/\tilde{\gamma}\sim\sqrt{\Gamma/\Delta_0}~p_F\lambda \simeq 1$. The
finite density of states $N(\omega\rightarrow0)/N_0 = \Gamma/\Delta_0 \sim
n_{imp}$, {\em linear} in the impurity concentration, is generated in the case
2) as well.

We now turn to the quasiparticle conductivity. We shall use the lowest
order bubble diagram with self-consistent Green functions with {\em no
vertex corrections} for this purpose, see for example~\cite{Lee}. For
the  dc  conductivity we get~\cite{Mahan}:
\begin{eqnarray}
\sigma(\omega\rightarrow0)=\frac{e^2}{\hbar}~
\frac{4 v^2_F}{\pi^2}\mathop{{\sum}'}_{\bbox{k}}\int
d\epsilon~(-\partial_{\epsilon}n(\epsilon))\times\nonumber\\
\times(|G''(\bbox{k},~\epsilon)|^2+|F''(\bbox{k},~\epsilon)|^2),
\label{cond}
\end{eqnarray}
where, linearizing the quasiparticle spectrum in the vicinity of the
nodes (see Fig.~\ref{fig1}),
$G''(\bbox{k},~\omega=0)=\gamma/(\gamma^2+(v_1k_1)^2+(v_Fk_3)^2),~F''(\bbox{k},~\omega
=0)=0$. The momentum integral in Eq.~(\ref{cond}) yields the final
formula Eq.~(\ref{sigv}) for $T=0$ with $O(T^2)$ corrections. This
formula holds for {\em any} model of disorder as long as
$v_1\lambda^{-1}/\gamma\ll 1$.

For the particular case of disorder considered in Eq.~(\ref{gamma2}),
the conductivity is
\begin{eqnarray}
\sigma(\omega\rightarrow0)=\frac{e^2}{\pi\hbar}~\frac{16}{\pi^3}~
\frac{\hbar}{m\lambda^2\Gamma}.
\end{eqnarray}
It is smaller than the normal state conductivity
$\sigma(\omega\rightarrow0)_{\rm normal}=(e^2/\pi\hbar)~(\epsilon_F/\Gamma)$
due to the small
factor $\hbar/(m\lambda^2\epsilon_F)\lesssim 1$ with $\lambda >a$. The
conductivity is also impurity-{\em dependent}, $\sigma\sim\Gamma^{-1}\sim
n^{-1}_{imp}$,
and for larger concentrations of impurities it decreases. This model
predicts that the  dc  conductivity at low temperatures should be
{\em inversly proportional to the scattering rate in the normal state and to
the impurity concentration}. We emphasize
that {\em both} a higher value of the conductivity in the
superconducting state {\em as well as} a strong impurity dependence of
the conductivity at low temperatures are observed experimentally in the
microwave absorption in YBCO~\cite{Bonn}. The conductivity increases
below the transition since the inelastic spin-fluctuation scattering
rate, contributing to the conductivity, is suppressed due to the
superconducting gap.

However  within this model the temperature dependence of
$\sigma\sim T$ at low temperatures remains unexplained, since we obtain
$T^2$ corrections at low enough temperature (see also~\cite{Doug2}). If
it is an intrinsic effect, it  indicates an important
physical effect, missing in this simple model.


We shall comment on the quasiparticle localization in the $d$-wave
superconductor. Originally the problem of the weak localization of
quasiparticles was considered for the $s$-wave
superconductivity~\cite{Ma}. For the case of  $d$-wave it was discussed
recently~\cite{Lee}.

Here we will show that the linearized Dirac-Bogoliubov equation in the {\em
 presence} of the impurity potential preserves time reversal
($\hat{T}$) symmetry and thus the weak localization in this state
belongs to the {\em orthogonal } universality class~\cite{Lee,Fradkin}. The
linearized Bogoliubov equation Eq.~(\ref{Dirac}) takes the form of a
Dirac equation with the scalar impurity potential $U(\bbox{r})$ playing
the role of the gauge potential. Although it appears that this
Hamiltonian violates $\hat{T}$, we will show that this is not the case.
The time reversal for electron operators is defined as
$\hat{T}c_{\alpha}^{}=\epsilon_{\alpha\beta}^{}c^{\dag}_{\beta},~\alpha,~\beta
=\uparrow\downarrow$, where $\epsilon_{\alpha\beta}^{}$ is the
antisymmetric tensor. Straightforward calculation yields for the time
inversion of the Nambu spinor $\Psi
=(c^{}_{\uparrow},c^{\dag}_{\downarrow})$
\begin{eqnarray}
\hat{T}\Psi =i\tau_2\Psi,~\hat{T}\Psi^{\dag}=\Psi^{\dag}(-i)\tau_2
\end{eqnarray}
The Dirac-Bogoliubov Hamiltonian transforms under $\hat{T}$ as follows $H =
\Psi^{\dag}((v_F\hat{k}_3 + U(r))\tau_3 +
v_1\hat{k}_1\tau_1)\Psi\stackrel{\hat{T}}{\longrightarrow}\Psi^{\dag}(-i\tau_2)
((v_F\hat{k}_3 + U(r))\tau_3 + v_1\hat{k}_1\tau_1)(i\tau_2)\Psi$ and

\begin{eqnarray}
H \stackrel{\hat{T}}{\longrightarrow}(-1)H.
\end{eqnarray}
Thus  the Dirac-Bogoliubov Hamiltonian transforms under time reversal
exactly as the energy, which changes sign under $\hat{T}$, and thus
time reversal symmetry is preserved. This fact also follows from the
observation that the original Bogoliubov Hamiltonian Eq.~(\ref{ham})
for a $d$-wave superconductor preserves the time reversal in the
presence of  scalar impurities. After linearization this symmetry
should be preserved as well. Thus the quasiparticle localization in the
impure $d$-wave superconductor belongs to the {\em orthogonal } class
of universality.

In conclusion, we find a new, {\em quasi-one-dimensional}, regime in
the superconducting scattering rate $\gamma$ and in the conductivity
$\sigma$ for the strong impurity scattering. The highly anisotropic
dispersion of the quasiparticle spectrum
$v_1/v_F\sim\Delta_0/\epsilon_F$ and finite range of the impurity
potential are {\em crucial } for this effect to take place. We argue
that the finite range $\lambda$ of the impurity potential is the result
of the strong antiferromagnetic correlations $\xi_{AFM}\sim 3a$ in the
normal phase. This effect might occur in {\em any} superconductors with
a linear quasiparticle spectrum in the nodes of the gap. We considered
a superconductor with $d_{x^2-y^2}$ symmetry of the gap and find that
for a reasonable scattering rate in the normal phase $\Gamma/T_c>0.2$, the {\em
quasi-one-dimensional } regime should occur. The
experimental consequence of this effect is that the conductivity at low
temperatures $\sigma\sim\Gamma^{-1}\sim n^{-1}_{imp}$ is impurity {\em
dependent} and is inversly proportional to the scattering rate in the
normal phase. We also show that even in the presence of scalar
impurities, time reversal symmetry in the $d$-wave superconductor is
preserved and argue that it leads to the weak quasiparticle
localization in the {\em orthogonal } universality class.

We are grateful to E. Abrahams, P.\ Lee, P.\ Littlewood, D.\ Pines and D.\ J.\
Scalapino for discussions. This work was supported by a J.\ R.\
Oppenheimer fellowship (A.\ B.), by the Department of Energy, and by
the Swedish Natural Science Research Council. Part of this work was
done at Aspen Center for Physics, whose support is also acknowledged.
A.\ R.\ would like to thank D.\ E.\ Peterson, J.\ L.\ Smith and K.\ S.\ Bedell
for their hospitality at the Los Alamos National Laboratory.



\begin{references}
\bibitem{gap}
We assume that the gap function $\Delta(\bbox{k})$ is not changed in
the presence of impurities, except for the suppressed amplitude
$\Delta_0$. Generally speaking we need to solve the complete Eliashberg
equations to investigate the effect of the impurities on the
superconducting state. One can show, however, that for impurities with
only $s$-wave phase shift the solution of the Eliashberg equations
gives the Green functions Eq.~(\ref{green}).
\bibitem{Born}
The fact that impurity scattering is {\em unitary} is crucial for this
{\em quasi-one-dimensional } limit to occur. Indeed the simple estimate
shows that for the {\em Born} limit with weak scattering
$v_1\lambda^{-1}/\gamma\gg 1$.
\end{references}
\end{document}